\documentclass[epj]{svjour}
\usepackage{graphics}
\begin{document}
\title{Baryon resonances and strong decays}
\author{T. Melde\inst{1} \and W. Plessas\inst{1}% etc
}                     % Do not remove
\institute{ Institute for Physics, Theoretical Physics, Graz, Austria
%\and the second here
}
\date{Received: date / Revised version: date}
\abstract{
Constituent quark models provide a reasonable description of the baryon 
mass spectra. However, even in the light- and strange-flavor sectors 
several intriguing shortcomings remain. Especially with regard to strong decays
of baryon resonances no consistent picture has so far emerged, and the
existing experimental data cannot be explained in a satisfactory manner. 
Recently first covariant calculations with modern constituent quark
models have become available for all $\pi$, $\eta$, and $K$ decay modes
of the low-lying light and strange baryons. They generally produced a remarkable
underestimation of the experimental data for partial decay widths. We summarize
the main results and discuss their impact on the classification of baryon
resonances into flavor multiplets. These findings are of particular relevance
for future efforts in the experimental investigation of baryon resonances.
 % end of PACS codes
} %end of abstract
\maketitle
\section{Introduction}
\label{intro}
Modern constituent quark models (CQMs) have reached an improved description
of the light and strange baryon spectra. Especially the masses of ground and
resonance states below about 2 GeV can generally be reproduced in a reasonable
manner~\cite{Plessas:2003av}. However, there are also some cases where disturbing discrepancies remain. A notable example is the $\Lambda(1405)$ resonance, which
cannot be explained by any CQM, relying only on three-quark configurations. In
addition there are considerable uncertainties above all in the $\Sigma$ and $\Xi$
excitation spectra due to a limited experimental data base.
 
Regarding strong decays of light and strange baryon resonances no satisfactory
description has yet been reached. So far, the mostly nonrelativistic CQM calculations
have produced results for partial decay widths that scatter around the experimental 
data~\cite{Stancu:1989iu,Capstick:1993th,Krassnigg:1999ky,Plessas:1999nb,Theussl:2000sj}.
In particular, it has proven difficult to explain some theoretical decay widths that
grossly overestimate the measured ones. Sometimes ad-hoc parametrizations (leading beyond
the CQMs) have been introduced in order to fit the data, but a consistent picture
has not emerged.

Recently, we performed covariant calculations of the $\pi$, $\eta$, and $K$ decay 
widths employing relativistic CQMs within the so-called point-form
spectator model (PFSM)~\cite{Melde:2005hy,Melde:2006yw,Sengl:2007yq}. The
direct predictions for partial decay widths produced a completely different pattern.
The relativistic results systematically underestimate the experimental data.
Nevertheless it became evident that relativity plays an immense
role. In particular it could be shown that a nonrelativistic reduction causes
large effects. They mainly result from truncations in the spin-coupling terms and the
neglect of Lorentz boosts thus explaining the variations in the nonrelativistic
calculations. Similar results for relativistic decay widths have been obtained by
the Bonn group using an instanton-induced CQM in the framework of the Bethe-Salpeter
equation~\cite{Metsch:2004qk,Migura:2007}.

Even though the relativistic decay calculations do not yet provide a
satisfactory explanation of the experimental decay widths, they produce a systematic
pattern of the results, which allows one to investigate existing ambiguities in
the classification of baryons into flavor multiplets. In the following we detail
some of the corresponding implications with regard to hyperon resonances.

\section{Systematics of strong decays}

From the comprehensive relativistic studies of $\pi$, $\eta$, and $K$ decay 
widths~\cite{Melde:2005hy,Melde:2006yw,Sengl:2007yq} a classification of
light and strange baryon resonances into $SU(3)$ flavor multiplets is
suggested as given in Tables~\ref{tab:multiplet_oct} and~\ref{tab:multiplet_decu}.
While in most instances the octet and decuplet assignments agree with the
ones by the PDG~\cite{PDBook}, there occur also some differences. The
$\Lambda(1810)$ is interpreted as flavor singlet, not as octet, and the $\Sigma(1620)$
falls into the octet involving the $N(1650)$. In addition, the $\Xi(1690)$ is
assigned to the octet involving $N(1440)$, the $\Sigma(1560)$ falls into the
octet involving $N(1535)$, the $\Sigma(1940)$ is assigned to the octet involving
$N(1700)$, and the $\Xi(1690)$ as well as $\Xi(1950)$ are members of the octets
involving $N(1440)$ and $N(1675)$, respectively. Except for the $\Lambda(1810)$
similar findings have also been obtained by Guzey and Polyakov~\cite{Guzey:2005vz}.
However, we remark that the $\Lambda(1810)$ is also identified as a flavor
singlet by Matagne and Stancu~\cite{Matagne:2006zf}.
In the context of our investigation we emphasize that the $\Sigma$
assignments for $J^P=\frac{1}{2}^-$ excitations are not completely safe, as
the pertinent resonances are not all sufficiently well established by experiment.

\begin{table}
\caption{
Classification of flavor octet baryons. The superscripts denote the percentages of
octet content in the mass eigenstates as calculated with the GBE
CQM~\cite{Glozman:1998ag}.
\label{tab:multiplet_oct}
}
{\begin{tabular}{p{1.0cm} p{1.4cm}p{1.4cm}p{1.4cm}p{1.4cm}}
\hline\noalign{\smallskip}
$(LS)J^P$&&&&\\
\noalign{\smallskip}\hline\noalign{\smallskip}
$(0\frac{1}{2})\frac{1}{2}^+$%
&$N(939)^{100}$%
&$\Lambda(1116)^{100}$%
&$\Sigma(1193)^{100}$% 
&$\Xi(1318)^{100}$%
\\
$(0\frac{1}{2})\frac{1}{2}^+$% 
&$N(1440)^{100}$%  
&$\Lambda(1600)^{96}$%
&$\Sigma(1660)^{100}$% 
&$\Xi (1690)^{100}$%
\\
$(0\frac{1}{2})\frac{1}{2}^+$%
&$N(1710)^{100}$%  
&%
&$\Sigma(1880)^{99}$% 
&% 
\\
$(1\frac{1}{2})\frac{1}{2}^-$%
&$N(1535)^{100}$%  
&$\Lambda(1670)^{72}$%
&$\Sigma(1560)^{94}$%
&%  
\\
$(1\frac{3}{2})\frac{1}{2}^-$%
&$N(1650)^{100}$% 
&$\Lambda(1800)^{100}$% 
&$\Sigma (1620)^{100}$%
&% 
\\
$(1\frac{1}{2})\frac{3}{2}^-$% 
&$N(1520)^{100}$%
&$\Lambda(1690)^{72}$% 
&$\Sigma(1670)^{94}$% 
&$\Xi(1820)^{97}$%
\\
$(1\frac{3}{2})\frac{3}{2}^-$%
&$N(1700)^{100}$% 
&%
&$\Sigma(1940)^{100}$% 
&% 
\\
$(1\frac{3}{2})\frac{5}{2}^-$% 
&$N(1675)^{100}$%  
&$\Lambda(1830)^{100}$%
&$\Sigma(1775)^{100}$% 
&$\Xi (1950)^{100}$%
\\
\noalign{\smallskip}\hline\noalign{\smallskip}
\end{tabular}}
\end{table}

\begin{table}
\begin{center}
\caption{
Classification of flavor decuplet baryons. The superscripts denote the percentages of
decuplet content in the mass eigenstates as calculated with the GBE
CQM~\cite{Glozman:1998ag}.
\label{tab:multiplet_decu}
}
\vspace{0.2cm}
{\begin{tabular}{p{1.0cm} p{1.4cm}p{1.4cm}p{1.4cm}p{1.4cm}}
\hline\noalign{\smallskip}
 $(LS)J^P$&&&&\\
\noalign{\smallskip}\hline\noalign{\smallskip}
$(0\frac{3}{2})\frac{3}2{}^+$
& $\Delta(1232)^{100}$ 
& $\Sigma(1385)^{100}$ 
& $\Xi(1530)^{100}$ 
& $\Omega(1672)^{100}$
\\
$(0\frac{3}{2})\frac{3}{2}^+$
& $\Delta(1600)^{100}$ 
& $\Sigma(1690)^{99}$
& & 
\\
$(1\frac{1}{2})\frac{1}{2}^-$
& $\Delta(1620)^{100}$  
&  $\Sigma (1750)^{94}$
& &
\\
$(1\frac{1}{2})\frac{3}{2}^-$
& $\Delta(1700)^{100}$ 
& & & 
\\
\noalign{\smallskip}\hline\noalign{\smallskip}\end{tabular}}
\end{center}
\end{table}

\begin{figure*}
\resizebox{\textwidth}{!}{%
\includegraphics{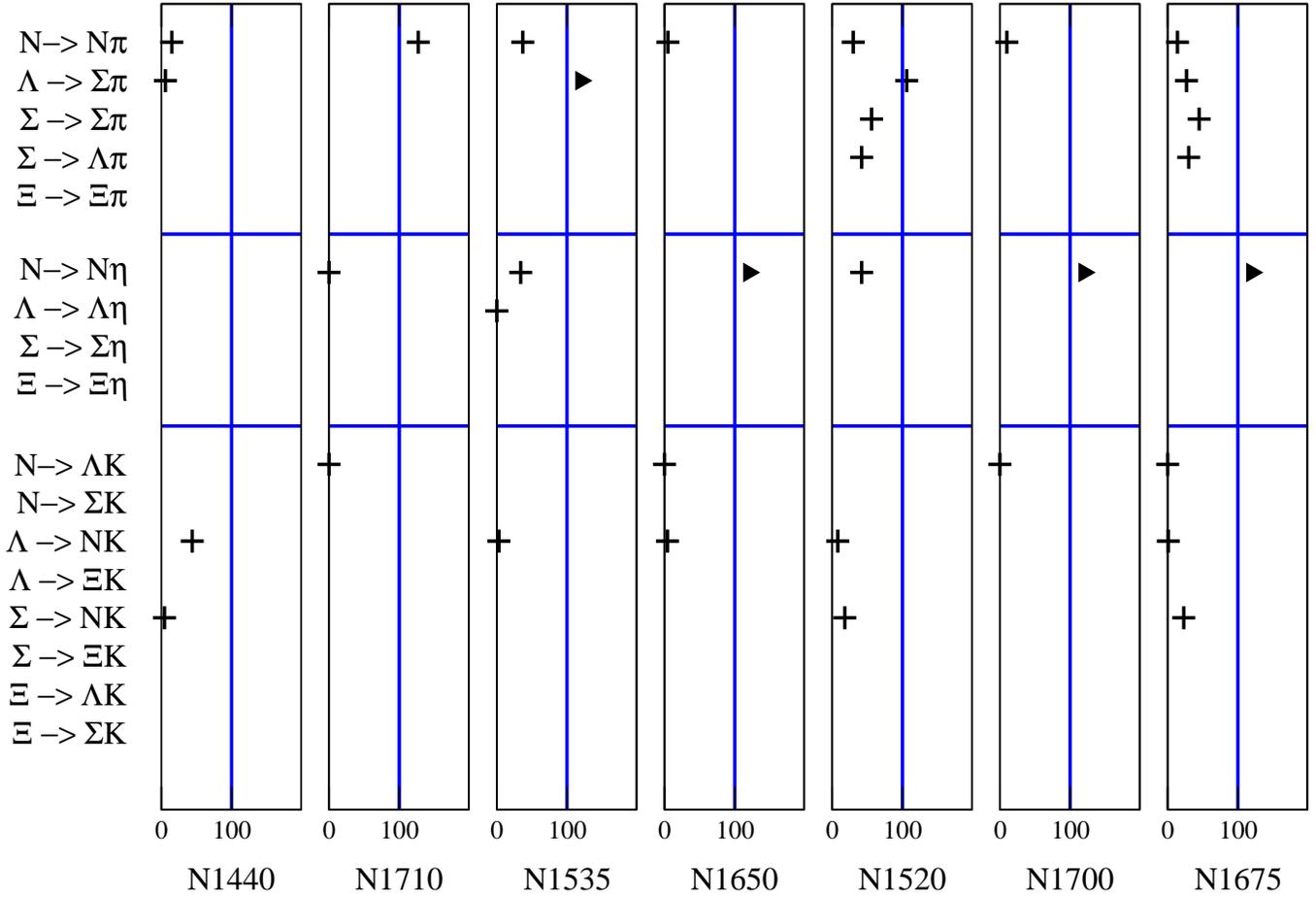}
}
\caption{Covariant predictions for $\pi$, $\eta$, and $K$ partial decay widths of octet
baryon resonances by the GBE CQM~\cite{Glozman:1998ag} according to
refs.~\cite{Melde:2005hy,Melde:2006yw,Sengl:2007yq}. Only established resonances (with
at least three-star status) are included. The theoretical results (marked by crosses)
are given as percentages of the best estimates for experimental widths reported by the
PDG~\cite{PDBook}. The (double) triangles point to results (far) outside the plotted
range.}
\label{fig:decay_graph_est}
\end{figure*}

Fig.~\ref{fig:decay_graph_est} shows the relativistic predictions of partial
decay widths for specific decay channels of octet baryon resonances. The results
are represented relative to the magnitudes of the experimental data. It is
immediately evident that in each one of the octets, denoted after the contained
nucleon member, the theoretical widths systematically underestimate the experiments
by similar amounts. There are only a few exceptions: The partial widths of
$\Lambda(1670) \rightarrow \Sigma \pi$ and $\Lambda(1690) \rightarrow \Sigma \pi$
come out too large; this can be explained by the large admixtures of flavour singlet contributions in each case (see Table~\ref{tab:multiplet_oct}).
Also the $N(1710)\rightarrow N\pi$ decay width appears to be unusually large;
the reason is not yet particularly clear, and it might be caused
by a deficiency in the theory and/or experiment (disagreeing partial wave analyses).
A further exception to underestimated widths consists in the $N(1650)\rightarrow N\eta$
decay, which represents a
notorious difficulty for CQMs. The remaining results in the $\eta$ channel should
not be taken too seriously as the corresponding phenomenological partial decay
widths are basically zero.

Of particular interest in the study of decay widths is the $J^P=\frac{1}{2}^-$
sector of the $\Sigma$ resonances. Here, the CQMs produce three lower-lying
eigenstates. From experiments only one established resonance (with at least
three-star) is reported, namely, the $\Sigma(1750)$. Its partial width for the
decay into $\Sigma \pi$ is quoted as being less than 8\% of the total width,
i.e. it should be rather small~\cite{PDBook}. On the other hand, the theoretical
decay widths of the lowest $J^P=\frac{1}{2}^-$ $\Sigma$ eigenstates turn out
to be relatively big~\cite{Melde:2006yw}. Certainly they would not fit
into the general pattern of Fig.~\ref{fig:decay_graph_est},
if they were interpreted as $\Sigma(1750)$. Now, the PDG also gives two more
$\Sigma$ resonances lying below the $\Sigma(1750)$, however, only with two-star
status. They are the $\Sigma(1620)$, whose $J^P$ is determined to be $\frac{1}{2}^-$,
and the $\Sigma(1560)$ without definite $J^P$ assignment. If one interprets the
two lower mass eigenstates as the $\Sigma(1560)$ and the $\Sigma(1620)$ and the
third eigenstate as the $\Sigma(1750)$, then a consistent pattern of the partial
decay widths is achieved~\cite{Melde:2006yw}. This naturally leads to the
assignments of the $\Sigma(1560)$ and the $\Sigma(1620)$ to the flavor octets
as given in Table~\ref{tab:multiplet_oct}. As a consequence the third eigenstate,
$\Sigma(1750)$, falls into the decuplet (cf. Table~\ref{tab:multiplet_decu}).
This classification is supported also by a more detailed investigation of
baryon resonance wave functions considering their specific spin-, flavor- and
spatial structures as resulting from relativistic CQMs.

\section{Conclusions and outlook}

Recent relativistic results for strong decay widths of baryon resonances
have produced a completely different pattern of CQM predictions. The magnitudes
of the various partial decay widths are generally too small and not compatible
with phenomenology. We have considered the direct predictions of CQMs without any
additional fitting of the results. The findings obtained within the PFSM
approach~\cite{Melde:2005hy,Melde:2006yw,Sengl:2007yq}
are surprisingly similar to the ones in the framework of the Bethe-Salpeter
equation~\cite{Metsch:2004qk,Migura:2007}. The defects hint to missing
contributions. One must bear in mind that the baryon ground and resonance states
are all described as bound three-quark eigenstates of the invariant mass operator.
Thus they have zero widths and cannot decay. The decay amplitudes are merely
calculated as transition matrix elements between bound states. Consequently
the shortcomings are not really surprising. Fitting the results a-posteriori
to the experimental data will not help in understanding the strong decays.
Rather one should think of improvements in the description of resonance states
and the decay operator.

The PFSM provides the simplest relativistic decay mechanism. It reduces to the
elementary emission model in the nonrelativistic limit~\cite{Melde:2006yw}.
However, it is not a
mere one-body operator but effectively includes many-body 
contributions~\cite{Melde:2004qu,Melde:2006jn}. Certainly the point-form
calculation is manifestly covariant and contains all relativistic effects
according to the spectator-model construction. Additional studies along this
line, employing different quark-meson couplings, could provide further insights.
Ultimately, however, improvements of the decay operator might be necessary
that go beyond the spectator model.

Another step towards improvements consists in an extension of CQMs to include
explicit couplings to the decay channels. In such a framework, the baryon states
will receive a finite width leading to a more realistic description of the excited
resonances. Of course, such a procedure will not
only have an effect on the widths, but also modify the (real) mass
values (cf., e.g., refs.~\cite{Gross:1992tj,Surya:1995ur}). As a result a
complete reconstruction of the CQMs will be necessary and the interpretation
of the theoretical baryon spectra might then appear in a different light.   

\begin{acknowledgement}
This work was supported by the Austrian Science Fund FWF, Project P19035.
Valuable contributions to this manuscript have been made by B. Sengl and
R.~F. Wagenbrunn.
\end{acknowledgement}

\end{document}